\documentclass[english]{article}
\usepackage[T1]{fontenc}
\usepackage[latin9]{inputenc}
\usepackage{amsmath}
\usepackage{amssymb}
\usepackage{graphicx}
\usepackage{esint}

\makeatletter

\providecommand{\tabularnewline}{\\}

\newcommand{\lyxaddress}[1]{
\par {\raggedright #1
\vspace{1.4em}
\noindent\par}
}

\makeatother

\usepackage{babel}
\begin{document}

\title{\textbf{\Large{}The role of the Havriliak-Negami relaxation in the
description of local structure of Kohlrausch's function in the frequency
domain. Part II}}

\author{{\normalsize{}J.S. Medina,}%
\thanks{{\footnotesize{}tlazcala@yahoo.es}%
}{\normalsize{} R. Prosmiti,$^{1}$ and J.V. Alemán$^{2}$}}

\maketitle

\lyxaddress{\emph{\footnotesize{}$^{1}$Instituto de Física Fundamental, IFF-CSIC,
Serrano 123, Madrid ES-28006, Spain}}

\lyxaddress{\emph{\footnotesize{}$^{2}$Departamento de Química, Facultad de
Ciencias del Mar, ULPGC, Campus Universitario de Tafira, Las Palmas
de G. Canaria ES-35017, Spain}}
\begin{abstract}
Two new sets of models for describing compactly the Fourier Transform
of Kohlrausch-Williams-Watts, both based on the \emph{adiabatic} variation
of parameters of a double Havriliak-Negami approximation along the
whole interval of frequencies, are presented. One of them is relying,
obviously, on the use of a well-behaved-pair of patches of the mentioned
type of approximants, $\mathcal{A}p_{2}HN(\omega)$. The other is
obtained by altering the simple functions $HN(\omega)$ and making
dissimilar the couple. They are proposed the guidelines of a new and
systematic approach with extended Havriliak-Negami functions which
is global, (non local), and of constant parameters. The latter at
the cost of a more complicated dependency with the low frequencies
than $1+(i\omega\tau_{HN})^{\alpha}$.
\end{abstract}

\section*{Introduction}

To the extent that the object of study of soft matter and fluids has
been passing from simple polar liquids to polymer, glasses and quasi-amorphous
materials, the phenomenology of rheological, or dielectric, relaxations
of physical systems has become increasingly complicated. And it is
not only that several types of those relaxations are superposed along
frequency space making difficult to distinguish among them, but that
employed functional form evolves from an easy one as Debye \cite{Deby 1913},
$\frac{1}{1+i\omega}$, to other more complex as Havriliak-Negami
\cite{Havr 1966,Havr 1967}, $\frac{1}{(1+(i\omega)^{\alpha})^{\gamma}}$,
after experimenting with intermediate stages as Cole-Cole \cite{Cole 1941},
$\frac{1}{1+(i\omega)^{\alpha}}$, and Cole-Davidson \cite{Davi 1951},
$\frac{1}{(1+i\omega)^{\gamma}}$.

Simultaneously something similar happens in time description while
we consider the different temporal scales implied, so certain habit
to model, --imposed by the mentioned physical phenomena and the discriminatory
ability of experimental equipment--, shows a methodological exhaustion.
In this sense any testing for the use of new relaxation functions,
giving account of the novel experimental records, is fully justified
\cite{Kahl 2010,Stan 2010}.

However, as has been quoted previously (\cite{Wint 1941,Humb 1945,Poll 1946,Will 1970,Will 1971,Lind 1980,Hilf 2002,Cape 2011}),
the swapping from one functional space to the other still continues
to be hard, often drawing upon, the researcher, efficient numerical
methods to perform such devious change \cite{Dish 1985,Scha 1996,Snyd 1999,Wutt 2009}.

Nevertheless it turns to be sometimes unsatisfactory this capability
for blind calculations as it does not provide many times of a general
view allowing for the interpretation and identification of these models
whose available information is fragmentary or incomplete. In this
sense a catalogue for formulae linking frequency and time realms \cite{Hilf 2002},
is a precious help while are appraised significant system parameters.
Besides it is also an essential partner of the numerical analysis
to bound errors and expose constructs \cite{Wutt 2009,Alva 1991,Alva 1993},
both characteristics of computer techniques.

We already have shown as a set of Weibull distributions \cite{Weib 1951},
($\beta t^{\beta-1}\exp-t^{\beta}$, $0<\beta\leq2$), in Fourier
space, $\psi_{\beta}(\omega)$, admit a good approximate description
by sums of Havriliak-Negami functions \cite{Macd 1986,Medi 2011,Medi 2015}.
Additionally in a precedent paper to this work \cite{Medi 2015},
it is established the local character of the approximation and how,
with slight variation of the parameters $\{\alpha_{1,2},\gamma_{1,2},\tau_{1,2},\lambda\}$
with frequency $\omega$, $Ap_{2}HN$ can describe a perfect fit with
the objective function, $\psi_{\beta}$. Such \emph{adiabatic} behavior
is commonly misunderstood as an argument against the approximation
by means of basic relaxation functions as Havriliak-Negami. This fact
it is best interpreted as the need for a wider family of relaxations
with a known local portrayal.

Thus in this job we will focus on taking advantage of such ``local''
information to build ``global'' functions that ameliorate the preceding
approximation in the whole range of frequencies, $[0,\infty)$. Also
the relative error of all proposed models in the present and previous
report \cite{Medi 2015} is depicted and tested against the real data
obtained from Fourier integrals.

\section{Uncommon approaches}

\subsection{\label{sub:The-Atlas}The global two-term approximant: version with
an atlas}

In short we have constructed two approximants of type \cite{Medi 2015},
\begin{equation}
\mathcal{A}p_{2}HN_{\alpha,\gamma,\tau,\lambda}(\omega)=\sum_{s=1}^{2}\frac{\lambda_{s}}{(1+(i\tau_{s}\omega)^{\alpha_{s}})^{\gamma_{s}}},\label{eq:1}
\end{equation}
($\lambda_{1}+\lambda_{2}=1$), for two different overlapping intervals
$\nu\in[0,500.0005]$ and $\nu\in[1.00,10^{12}]$, (or $\nu\in[1.00,10^{7}]$
if $\beta>1$). Besides if we consider how the relative error between
moduli of approximant and function behaves as frequency varies, (\emph{i.e.
}it stabilizes at an almost constant value never greater than 0.2\%
for high frequencies and $0<\beta\leq2$), the upper bound of second
interval can be extended without a big amount of error to an unlimited
frequency. It means that there are two charts $(\mathcal{A}p_{2,l}HN(\omega),\Omega_{l})$
and $(\mathcal{A}p_{2,h}HN(\omega),\Omega_{h}^{*})$ with $\Omega_{l}\equiv2\pi\times$
$(0,500.0005)$ and $\Omega_{h}^{*}\equiv2\pi\times$ $(1,\infty)$
that reconstruct in an acceptable way the function $\psi_{\beta}(\omega)$
in the whole interval $(0,\infty)$, plus the value at $\psi_{\beta}(0)=1$
as an imposed condition.

Now, all that is needed to obtain a global solution is a way to merge
both charts without overlaps, following a standard procedure. We will
resort to an smooth, and monotonously increasing, function defined
as:%
\footnote{A practical production of function $\mathcal{W}_{i,s}(\omega)$ can
be made following instructions found in ref. \cite{Warn 1971}, lemma
1.10, p. 10.%
}
\[
\mathcal{W}_{i,s}(\omega)=\begin{cases}
\begin{array}{cc}
0 & :\\
(0,1) & :\\
1 & :
\end{array} & \begin{array}{c}
\omega\leq\omega_{i}\\
\omega_{i}<\omega<\omega_{s}\\
\omega_{s}\leq\omega
\end{array}\end{cases},
\]
with $\omega_{i}$ and $\omega_{s}$ chosen arbitrarily. So starting
from locally adjusted functions which are properly selected it is
possible to write a suitable approximation to function $\psi_{\beta}$,
in the whole interval $[0,\infty)$, as:
\begin{equation}
\mathcal{A}p_{2}HN(\omega)=\mathcal{A}p_{2,l}HN(\omega)\times(1-\mathcal{W}_{i,s})+\mathcal{A}p_{2,h}HN(\omega)\times\mathcal{W}_{i,s}.\label{eq: 2}
\end{equation}
With our particular choice $\omega_{i}=2\pi$ and $\omega_{s}=4\pi$,
(or $\omega_{s}=2.2\pi$ if $\beta>1$), we have finally laid down
a global surrogate of Havriliak-Negami type for the Fourier Transform
of Weibull function $\psi_{\beta}(\omega)$, whenever $0<\beta\leq2$.

\subsection{\label{sub:The-global-two-term}The global two-term approximant:
version with adiabatic parameters}

\subsubsection{The stretched case $\beta<1$}

While the chart for low frequencies is a quite good approximation
to function $\psi_{\beta}(\omega)$ it is not however a perfect matching
in the mentioned range. Thus, there still is room for improving the
fit as the difference reaches a peak of 1.3\% for some of the lower
frequencies, $\omega<10\delta\omega$, at medium values of beta, \emph{i.e.}
$\beta\approx0.30$, and working with an $r_{2}$ sampling, ($\delta\omega=1/999.999$).
So the logical next step would be to add another Havriliak-Negami
term to the approximant to fill the gap, although using some restrictions,
(\emph{v. gr. }over $\alpha\cdot\gamma$ products), to avoid proliferation
of parameters and to hold the resemblance of a truncated series. The
procedure works, however one should cope with other problems of the
multi-parameter optimization as the non isolated loci of minima, the
multiplicity of them, and the competition among coefficients for some
regions, (which prevents a broad and proper allocation of them in
the search space). It is then a good idea to analyze why a small gap
befalls precisely at very low frequencies in order to correct it or
design a strategy for the restrictions of new incoming terms of a
series, or an approximant.
\begin{table}
\centering{}\emph{\footnotesize{}}%
\begin{tabular}{ccc}
\hline 
\emph{\tiny{}Formulae} & {\tiny{}$\beta A\exp[\sum_{s=1}^{5}a_{s}(1-\beta)^{s}]$} & {\tiny{}$\exp[-(\frac{B}{1-\beta+\epsilon})^{2}]+\sum_{s=1}^{4}d_{s}(1-\beta)^{s}$}\tabularnewline
\hline 
\hline 
{\tiny{}}%
\begin{tabular}{c}
{\footnotesize{}$\frac{Parameters}{Constants}$}\tabularnewline
{\footnotesize{}$A$}\tabularnewline
{\footnotesize{}$a_{1}$}\tabularnewline
{\footnotesize{}$a_{2}$}\tabularnewline
{\footnotesize{}$a_{3}$}\tabularnewline
{\footnotesize{}$a_{4}$}\tabularnewline
{\footnotesize{}$a_{5}$}\tabularnewline
{\footnotesize{}Corr.}\tabularnewline
\end{tabular} & {\footnotesize{}}%
\begin{tabular}{c|c}
\multicolumn{1}{c}{{\footnotesize{}$\alpha_{2}$}} & {\footnotesize{}$\hat{\alpha}_{3}$}\tabularnewline
{\footnotesize{}0.886396} & {\footnotesize{}1.55239}\tabularnewline
{\footnotesize{}0.606292} & {\footnotesize{}0.282576}\tabularnewline
{\footnotesize{}1.46112} & {\footnotesize{}6.08038}\tabularnewline
{\footnotesize{}-8.94199} & {\footnotesize{}-32.0705}\tabularnewline
{\footnotesize{}13.2755} & {\footnotesize{}47.4824}\tabularnewline
{\footnotesize{}-6.3231} & {\footnotesize{}-22.507}\tabularnewline
\hline 
\multicolumn{1}{c}{\emph{\footnotesize{}0.999970}} & \emph{\footnotesize{}0.999913}\tabularnewline
\end{tabular} & {\footnotesize{}}%
\begin{tabular}{c|c}
\multicolumn{1}{c}{{\footnotesize{}$\frac{Parameter}{Constants}$}} & {\footnotesize{}$1-\lambda$}\tabularnewline
{\footnotesize{}$B$} & {\footnotesize{}0.186694}\tabularnewline
{\footnotesize{}$d_{1}$} & {\footnotesize{}1.65185}\tabularnewline
{\footnotesize{}$d_{2}$} & {\footnotesize{}-4.51252}\tabularnewline
{\footnotesize{}$d_{3}$} & {\footnotesize{}3.81826}\tabularnewline
{\footnotesize{}$d_{4}$} & {\footnotesize{}-0.916873}\tabularnewline
\hline 
\multicolumn{1}{c}{{\footnotesize{}Corr.}} & \emph{\footnotesize{}0.996672}\tabularnewline
\end{tabular}\tabularnewline
\hline 
\end{tabular}\protect\caption{\emph{\footnotesize{}\label{tab:Tab Ia-II}Optimization parameters
of the Eq. \ref{eq: 3} approximant, }{\footnotesize{}(}\emph{\footnotesize{}case
$\beta\leq1$}{\footnotesize{}): Formulas, (with $\epsilon=10^{-180}$),
and their constants for $\alpha_{2}$, $\hat{\alpha}_{3}$ and $1-\lambda$.}}
\end{table}

The first derivative of $\psi_{\beta}(\omega)$ $=\int_{0}^{\infty}\beta t^{\beta-1}e^{-t^{\beta}-i\omega t}\mathrm{d}t$
with respect the angular frequency is written as:
\[
\lim_{\omega\rightarrow0^{+}}\frac{\mathrm{d}\psi_{\beta}(\omega)}{\mathrm{d}\omega}=\lim_{\omega\rightarrow0^{+}}-i\int_{0}^{\infty}\beta t^{\beta}e^{-t^{\beta}-i\omega t}\mathrm{d}t=-i\int_{0}^{\infty}\beta t^{\beta}e^{-t^{\beta}}\mathrm{d}t=-i\Gamma(\frac{1}{\beta}+1),
\]
while the first derivative for the Havriliak-Negami function is expressed
as:
\[
\frac{\mathrm{d}}{\mathrm{d}\omega}\frac{1}{(1+(i\omega\tau)^{\alpha})^{\gamma}}=\frac{-\gamma\alpha(i\tau)^{\alpha}\omega^{\alpha-1}}{(1+(i\omega\tau)^{\alpha})^{\gamma+1}}\longrightarrow_{\omega\rightarrow0^{+}}\begin{cases}
\begin{array}{cc}
0 & :\\
-i\gamma\tau & :\\
-i^{\alpha}\infty & :
\end{array} & \begin{array}{c}
\alpha>1\\
\alpha=1\\
\alpha<1
\end{array}\end{cases},
\]
which implies, as $-i^{\alpha}=$ $-\cos\frac{\pi\alpha}{2}-i\sin\frac{\pi\alpha}{2}$
takes values in the third quadrant, that no lineal combination of
two vectors 
\[
(\frac{\mathrm{d}HN_{1}(\omega)}{\text{\ensuremath{\mathrm{d}\omega}}}/\omega^{\alpha_{1}-1})_{\omega=0}\mathrm{\quad and}\quad(\frac{\mathrm{d}HN_{2}(\omega)}{\text{\ensuremath{\mathrm{d}\omega}}}/\omega^{\alpha_{2}-1})_{\omega=0}
\]
can give a vector parallel to $-i$ when both $\alpha_{1,2}<1$. Nor
it is possible with only one $\alpha<1$. In such a situation, \emph{i.e.}
$\alpha_{1}<1$ or $\alpha_{2}<1$, the magnitude of at least one
modulus will be infinity due to factor $\omega^{\alpha_{1,2}-1}$
what gives a reason, --the momentary diminishing of any $|HN_{1,2}(\omega)|$
in the vicinity of $\omega\approx0^{+}$ is quicker than that of $|\psi(\omega)|$--,
for the underestimation of $\psi_{\beta}$ by $\mathcal{A}p_{2}HN_{\alpha,\gamma,\tau,\lambda}$
in the range of low frequencies.
\begin{table}
\centering{}{\tiny{}}%
\begin{tabular}{ccc}
\hline 
{\tiny{}$(1-\beta)\big\{\frac{1}{\sum_{r=0}^{2}b_{r}\beta^{r}}+\sum_{s=1}^{2}C_{s}\cos(\zeta_{s}\beta+\phi_{s})\big\}$} & {\tiny{}$\exp\big[(\frac{1}{\beta})^{2}\sum_{s=0}^{6}c_{s}(1-\beta)^{s}\big]$} & {\tiny{}$\frac{\exp[-M\beta^{d}]}{\sum_{r=0}^{3}b_{r}\beta^{r}}$}\tabularnewline
\hline 
\hline 
{\footnotesize{}}%
\begin{tabular}{c|c}
\multicolumn{1}{c}{{\footnotesize{}$\frac{Parameter}{Constants}$}} & {\footnotesize{}$\log_{10}\tau_{1}$}\tabularnewline
{\footnotesize{}$b_{0}$} & {\footnotesize{}0.0979577}\tabularnewline
{\footnotesize{}$b_{1}$} & {\footnotesize{}0.757724}\tabularnewline
{\footnotesize{}$b_{2}$} & {\footnotesize{}-0.228709}\tabularnewline
{\footnotesize{}$C_{1}$} & {\footnotesize{}-0.142877}\tabularnewline
{\footnotesize{}$\zeta_{1}$} & {\footnotesize{}22.2049}\tabularnewline
{\footnotesize{}$\phi_{1}$} & {\footnotesize{}-7.14011}\tabularnewline
{\footnotesize{}$C_{2}$} & {\footnotesize{}0.649622}\tabularnewline
{\footnotesize{}$\zeta_{2}$} & {\footnotesize{}-8.86238}\tabularnewline
{\footnotesize{}$\phi_{2}$} & {\footnotesize{}5.19663}\tabularnewline
\hline 
\multicolumn{1}{c}{{\footnotesize{}Corr.}} & \emph{\footnotesize{}0.999836}\tabularnewline
\end{tabular} & {\footnotesize{}}%
\begin{tabular}{c|c}
\multicolumn{1}{c}{{\footnotesize{}$\frac{Parameter}{Constants}$}} & {\footnotesize{}$\tau_{2}$}\tabularnewline
{\footnotesize{}$c_{0}$} & {\footnotesize{}-0.796921}\tabularnewline
{\footnotesize{}$c_{1}$} & {\footnotesize{}4.14764}\tabularnewline
{\footnotesize{}$c_{2}$} & {\footnotesize{}-2.98386}\tabularnewline
{\footnotesize{}$c_{3}$} & {\footnotesize{}-14.9068}\tabularnewline
{\footnotesize{}$c_{4}$} & {\footnotesize{}33.8185}\tabularnewline
{\footnotesize{}$c_{5}$} & {\footnotesize{}-26.9545}\tabularnewline
{\footnotesize{}$c_{6}$} & {\footnotesize{}7.67633}\tabularnewline
\hline 
\multicolumn{1}{c}{{\footnotesize{}Corr.}} & \emph{\footnotesize{}0.995682}\tabularnewline
\end{tabular} & {\footnotesize{}}%
\begin{tabular}{c|c}
\multicolumn{1}{c}{{\footnotesize{}$\frac{Parameter}{Constants}$}} & {\footnotesize{}$\log_{10}\tau_{3}$}\tabularnewline
{\footnotesize{}$M$} & {\footnotesize{}2.74977}\tabularnewline
{\footnotesize{}$d$} & {\footnotesize{}4.40586}\tabularnewline
{\footnotesize{}$b_{0}$} & {\footnotesize{}-9.15927$\times10^{-6}$}\tabularnewline
{\footnotesize{}$b_{1}$} & {\footnotesize{}0.559288}\tabularnewline
{\footnotesize{}$b_{2}$} & {\footnotesize{}0.159154}\tabularnewline
{\footnotesize{}$b_{3}$} & {\footnotesize{}-0.641584}\tabularnewline
\hline 
\multicolumn{1}{c}{{\footnotesize{}Corr.}} & \emph{\footnotesize{}0.999992}\tabularnewline
\end{tabular}\tabularnewline
\hline 
\end{tabular}\protect\caption{\emph{\footnotesize{}\label{tab:Tab Ib-II}Optimization parameters
of the Eq. \ref{eq: 3} approximant, }{\footnotesize{}(}\emph{\footnotesize{}case
$\beta\leq1$}{\footnotesize{}): Formulas and their constants for
$\tau_{1}$, $\tau_{2}$ and $\tau_{3}$.}}
\end{table}

All this confront us with the fact of finding alternatives to $\alpha_{1,2}<1$,
and under the conditions of module and direction of the derivative
of $\psi_{\beta}$ these are simply $\alpha_{1,2}=1$, or $\alpha_{1}=1$
and $\alpha_{2}>1$, (or the converse pair of indices). That contradicts
the empirical finding for parameters $\alpha_{1,2}$ we made with
the $r_{2}$ sampling in the range of medium to low frequencies which
supports heavily the condition $\alpha_{1,2}<1$ when $\beta<0.60$
and the conditions $\alpha_{1}<1$ and $\alpha_{2}\sim1$ when $\beta>0.60$.
Consequently the double approximant will always hold this gap in the
region of small frequencies unless of course we add some local modification
to parameters $\{\alpha,\gamma,\tau,\lambda\}_{1,2}$ near $\omega\approx0^{+}$.

In summary there are three zones in the $\omega$-space where the
coefficients $\{\alpha,\gamma,\tau,\lambda\}_{1,2}$ are similar in
magnitude, --comparing equal symbols--, although with a different
behaviour as functions of $\beta$. And they change from one to another
conduct when they are shifted across the intervals of frequency. This
means the set of parameters depends on $\omega$, \emph{i.e. }$\{\alpha,\gamma,\tau,\lambda\}_{1,2}=\{\alpha,\gamma,\tau,\lambda\}_{1,2}(\omega)$,
but they are of a very slow variation through the whole interval of
frequencies $[0,\infty)$. They are \emph{adiabatic} coefficients
of the approximant $\mathcal{A}p_{2}HN_{\alpha,\gamma,\tau,\lambda}(\omega)$.
\begin{figure}
\centering{}\includegraphics[width=0.8\columnwidth]{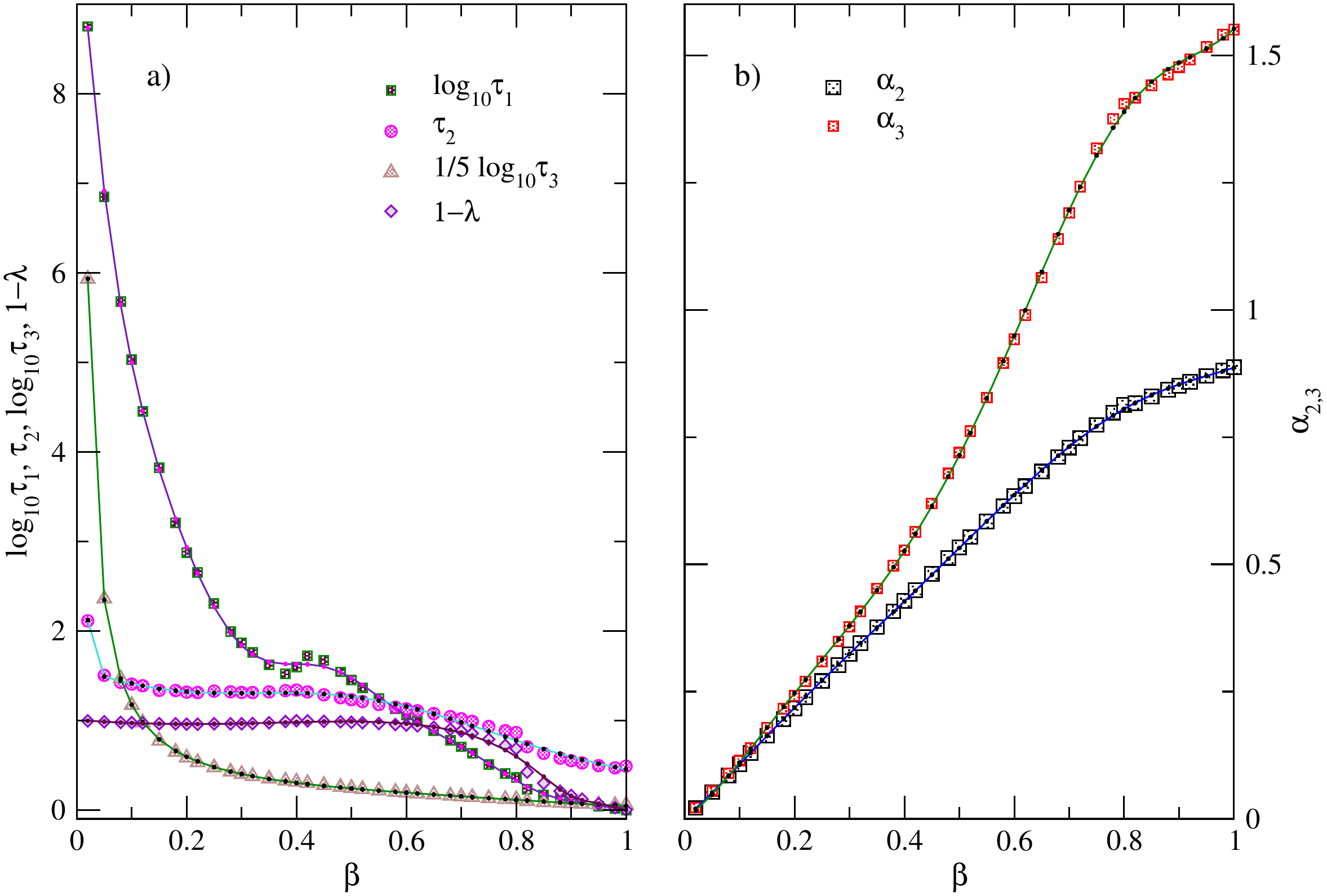}\protect\caption{{\footnotesize{}\label{Figure 1-II}Optimization parameters of modified
Havriliak-Negami approximant that is shown in Eq. \ref{eq: 3}. They,
$\alpha_{2}$, $\tau_{1}$, $\tau_{2}$ and $1-\lambda$, follow strictly
Eq. \ref{eq: 3} while $\tau_{3}$ and $\alpha_{3}$, ($\hat{\alpha}_{3}$
in text), are as in Eq. \ref{eq: 4}, the latter a simple estimation
for the }\emph{\footnotesize{}mollifier}{\footnotesize{}. $\beta\leq1$.
Solid lines are mathematical adjustments for parameters, which are
given in tables \ref{tab:Tab Ia-II} and \ref{tab:Tab Ib-II}.}}
\end{figure}

Therefore we are looking for a function reproducing the main traits
of $\psi_{\beta}(\omega)$, namely: $(\frac{\mathrm{d\psi_{\text{\ensuremath{\beta}}}}}{\mathrm{d}\omega})_{\omega=0}\parallel-i$,
$\omega^{\beta}*|\psi_{\beta}(\omega)|\stackrel{\omega\rightarrow\infty}{\longrightarrow}\mathcal{O}(1)$,
and $\psi_{\beta}(\omega)$ $\approx\mathcal{O}(\mathcal{A}p_{2}HN(\omega))$
locally $\forall\omega\in[0,\infty)$ with adiabatic coefficients
as parameters in the fit. So with that in mind, and respecting conditions
$(\alpha\cdot\gamma)_{1,2}=\beta$ for $\beta<1$, our candidate will
take the form:

\begin{equation}
\psi_{\text{\ensuremath{\beta}}}(\omega)\cong\mathcal{AM}_{l,2}HN(\omega)\equiv\frac{\lambda}{(1+i\omega\tau_{1})^{\beta}}+\frac{1-\lambda}{(1+\mathcal{M}_{l}(\omega)(i\omega\tau_{2})^{\alpha_{2}})^{\frac{\beta}{\alpha_{2}}}},\label{eq: 3}
\end{equation}
with $\mathcal{M}_{l}(\omega)$, the \emph{mollifier} of the Havriliak-Negami
function, verifying the following boundary conditions 
\[
\lim_{\omega\rightarrow\infty}\mathcal{M}_{l}(\omega)=1\quad\mathrm{and}\quad\mathcal{M}_{l}(\omega)\approx(\omega\tau_{3})^{\alpha_{3}},\; if\;\omega\approx0,
\]
where $\alpha_{2}+\alpha_{3}$ $\geq1$.

Undoubtedly to determine the mollifier it is not an easy task and
surely its expression as a series could be as difficult of obtaining
as the one of $\psi_{\beta}$, and yet the conditions we have imposed
on $\mathcal{M}_{l}(\omega)$ may allow an easy estimate of it. So
we will put forward as an estimator of mollifier$\mathcal{M}_{l}(\omega)$
the function
\begin{equation}
\hat{\mathcal{M}_{l}}(\omega)\equiv[\frac{2}{\pi}\arctan((\omega\tau_{3})^{\hat{\text{\ensuremath{\alpha}}}_{3}})]^{N},\label{eq: 4}
\end{equation}
setting $N=3$ as an appropriate average after a timely optimization
for some values of $\beta$.

\subsubsection{The squeezed case $\beta>1$}

As $\alpha_{1}>1$ and $\alpha_{2}<1$ for $2\geq\beta>1$, when the
adjustment to $r_{2}$ sampling is done, it is arguable that a lineal
combination of both direction vectors, $(\frac{\mathrm{d}HN_{1,2}(\omega)}{\text{\ensuremath{\mathrm{d}\omega}}}/\omega^{\alpha_{1,2}-1})$,
at $\omega=0$ can be parallel to $-i$, since they take values in
opposite quadrants (\emph{i.e. }second and third ones). Nevertheless
the magnitude of the derivatives, (zero and infinity respectively),
avoids such result, and again the admissible options for $\alpha_{1,2}$
are those of the case $\beta<1$ which as we pointed out contradict
the numerical findings. We must proceed again with some modifications
of the approximant, however it is not possible now to extend the model
of equation \ref{eq: 3} to this situation. Firstly the logarithmic
second derivatives of $\log_{10}|\psi_{\beta}|$ are qualitatively
different for $\beta<1$ and $\beta>1$ cases. (See figures 1 and
2 in \cite{Medi 2015}). Second only the $\alpha_{2}\cdot\gamma_{2}=\beta$
condition holds while the other becomes $\alpha_{1}\cdot\gamma_{1}\simeq2\beta$
for the asymptotic behaviour of tails. Besides in the low to medium
frequencies range the products $(\alpha\cdot\gamma)_{1,2}$ present
a nonlinear trend greater in many cases than the corresponding linear
condition of high frequency. (See figure 9 in \cite{Medi 2015}).
In consequence we introduce a peculiar model with two characteristic
times in the first Havriliak-Negami relaxation which jointly with
the two terms structure of the whole approximant will retain the mentioned
characteristics and restrictions for tails, with the sole exception
of $\alpha_{0}\cdot\gamma_{1}$ which is to be determined by fitting.
The latter in light of the numerical results appears a spurious property
or at least a virtual one conceived to explain the sudden change of
curvature in a small interval of frequencies, $\nu\lesssim1$.
\begin{figure}
\centering{}\includegraphics[width=0.8\columnwidth]{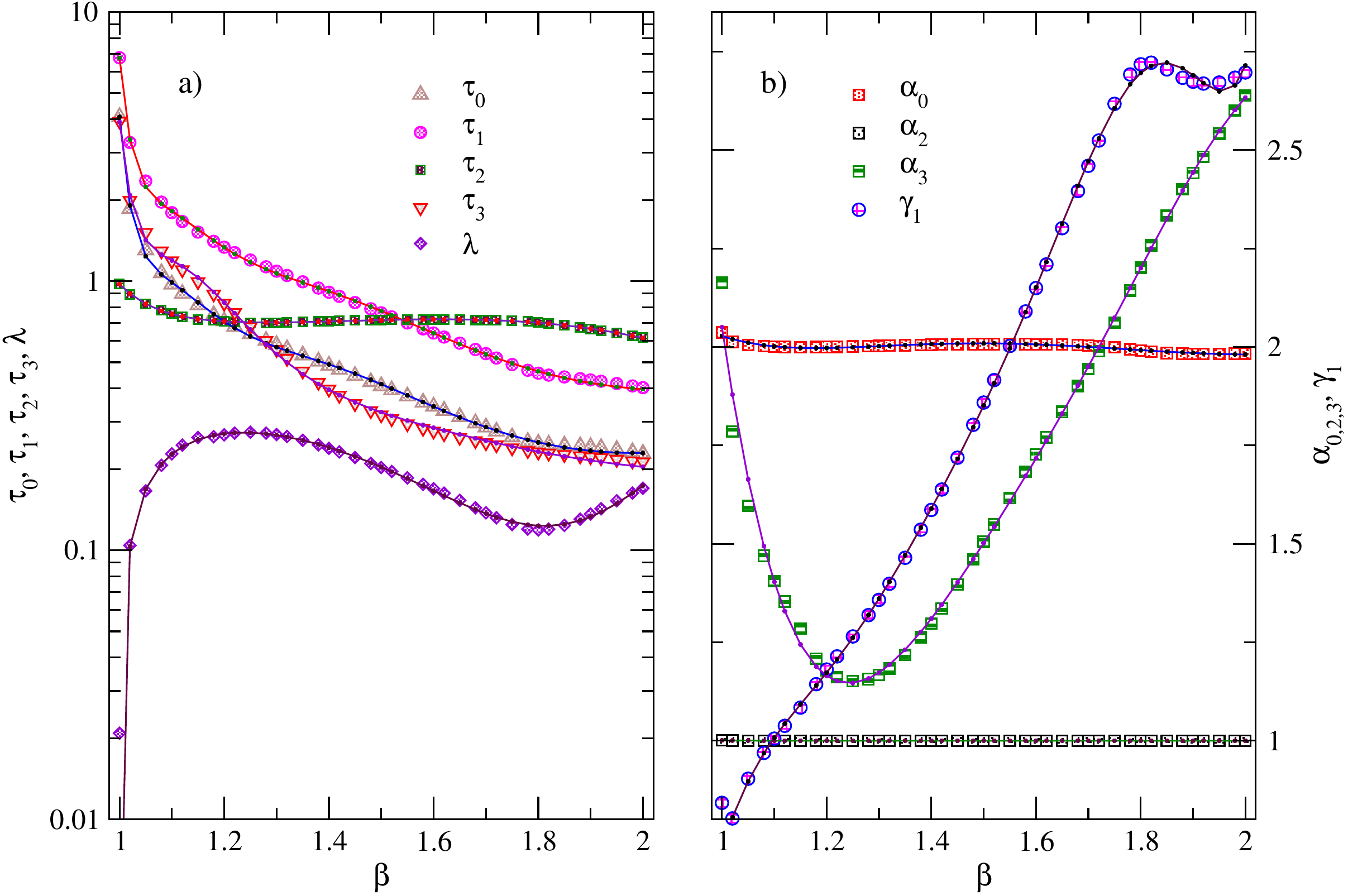}\protect\caption{{\footnotesize{}\label{Figure 2-II}For $\beta\geq1$ an approximation
to $\psi_{\beta}$ is done using the modified approximant $\mathcal{AM}_{g,2}HN$
set in Eq. \ref{eq: 5}. At left panel parameters: $\text{\ensuremath{\tau}}_{0}$,
$\tau_{1}$, $\tau_{2}$, $\tau_{3}$ and $\lambda$. At right panel:
$\alpha_{0}$, $\alpha_{2}$, $\alpha_{3}$ and $\gamma_{1}$. Except
for $\tau_{3}$ and $\alpha_{3}$, ($\hat{\alpha}_{3}$ in text),
all of them are defined in Eq. \ref{eq: 5}. The former are established
in Eq. \ref{eq: 6} which describes an estimator of the current }\emph{\footnotesize{}mollifier}{\footnotesize{}.
As in previous graphics the parameters are adjusted with suitable
mathematical expressions, (solid lines), given in tables \ref{tab:Tab IIa-II}
and \ref{tab:Tab IIb-II}.}}
\end{figure}

Thus when $\beta>1$ the global formula for the approximant it reads:
\begin{equation}
\psi_{\text{\ensuremath{\beta}}}(\omega)\cong\mathcal{AM}_{g,2}HN(\omega)\equiv\frac{\lambda}{(1+i\omega\tau_{1}+(i\omega\tau_{0})^{\alpha_{0}})^{\gamma_{1}}}+\frac{1-\lambda}{(1+\mathcal{M}_{g}(\omega)(i\omega\tau_{2})^{\alpha_{2}})^{\frac{\beta}{\alpha_{2}}}},\label{eq: 5}
\end{equation}
with the boundary conditions for the mollifier as
\[
\lim_{\omega\rightarrow\infty}\mathcal{M}_{g}(\omega)=1\quad\mathrm{and}\quad\mathcal{M}_{g}(\omega)\sim\mathcal{O}(1),\; if\;\omega\approx0,
\]
where $\alpha_{0}>1$ and $\alpha_{2}\gtrsim1$.

Again an exact expression for the mollifier is out of scope of present
study and we settle for an estimator like
\begin{equation}
\hat{\mathcal{M}}_{g}(\omega)\equiv[\frac{1}{\sqrt[n]{2}}\{1+(\sqrt[n]{2}-1)\frac{2}{\pi}\arctan(\omega\tau_{3})^{\hat{\alpha}_{3}}\}]^{n},\label{eq: 6}
\end{equation}
and with an \emph{ad hoc }choice of $n=3$. With such estimator we
will obtain a good approximation to $|\psi_{\beta}|$ which deviates
slightly in a neighborhood of $\omega\simeq2\pi$, (the zone where
the maximum of curvature happens in logarithmic scale), although it
describes fairly the body of function and quite well the trend and
values of tail. (See figure 2 in \cite{Medi 2015}).

\subsubsection{Analysing trends of parameter graphics}

Finally in figures \ref{Figure 1-II} and \ref{Figure 2-II} we display
the parameters of estimated functions $\hat{\mathcal{AM}_{l,2}HN(\omega)}$
and $\hat{\mathcal{AM}_{g,2}HN(\omega)}$ and their adjustments as
functions of $\beta$ (respectively $\beta<1$ and $\beta>1$). In
this occasion the interval of frequencies is up to $\nu=10^{12}$,
and up to $\nu=10^{6}$ depending upon choice of $\beta$, and the
sampling of frequencies is what we called logarithmically homogeneous,
\emph{i.e. }$r_{sl}$. In both cases all the curves have a break,
or turnaround, more or less evident according to each one. This occurs
for each curve, --within same $\beta$ case--, in the same point ($\beta\approx0.80$
and $\text{\ensuremath{\beta}}\approx1.80$). Also during the course
of several optimizations such points have changed marginally and the
breaks have increased or diminished their sharpness according as we
changed the average exponent ($N$, $n$) of equations \ref{eq: 4}
and \ref{eq: 6}, data weights or samplings ($r_{2}$, $r_{sl}$).
So in conclusion we interpret that such abnormalities are a result
of the shape of estimators.

\subsubsection*{Stretched instance, $\beta<1$}

The best option in case $\text{\ensuremath{\beta}}<1$ is to weigh,
--while using \texttt{xmgrace} to get a fit \cite{Turn 1998}--, the
tails with option $1/Y^{2}$ to soften the jump and obtain an even
adjustment all the way in the interval of frequencies. The value of
$N$ also could be lowered but the price to pay is an increasing error
for all the matching between both functions, (approximant and $\psi_{\beta}$),
around values of $\beta\in(0.1,0.3)$ and $\beta\in(0.7,0.8)$. On
the other hand the ability of the new function $\mathcal{AM}_{l,2}HN(\omega)$
for describing the effect that slow variation parameters $\{\alpha,\gamma,\tau\}_{1,2}$
would have in the original Havriliak-Negami functions fully justifies
the introduction of mollifier $\mathcal{M}_{l}(\omega)$. Unfortunately
the expression of its estimator does not seems good enough in the
vicinity of $\omega\approx0$ since the results do not fulfill the
required condition $\alpha_{2}+N\hat{\alpha}_{3}\geq1$ at all when
$\beta\rightarrow0^{+}$. This is a consequence of having frozen the
exponent at $N=3$, we should increase its value till infinity to
compensate the empirical trends of $\alpha_{2}$ and $\hat{\alpha}_{3}$
to be zero when $\beta\rightarrow0^{+}$. However the first term of
$\hat{\mathcal{AM}_{l,2}HN(\omega)}$, an almost residual one since
$\lambda\approx0$ for $\beta<0.8$, seems to balance numerically
this mathematical unsuitability of the second term of the approximant
in the description of $\psi_{\beta}(\omega\approx0)$. And that is
possible since there is no conflict in accounting for a slow diminishing
$|\psi_{\beta}|$ in the neighborhood of $\omega\approx0$, ($\beta<0.10$),
using a fast decaying Havriliak-Negami type function, (\emph{i.e.}
of large $\tau$), with the sampling step that we used. For such small
values of $\beta$ a neighborhood of zero where $|\psi_{\beta}|\sim1$
is so elusive that a frequency step of $\delta\nu=10^{-8}$ is too
large for considering a description of the modulus gradual decay.
(In tables \ref{tab:Tab Ia-II} and \ref{tab:Tab Ib-II} we wrote
the mathematical expressions for the six parameters of $\hat{\mathcal{AM}_{l,2}HN(\omega)}$
as curves depending of variable $\beta$).
\begin{table}
\begin{centering}
\begin{tabular}{cccc}
\hline 
\emph{\tiny{}Formulae} & {\tiny{}$\exp\big[\{\sum_{s=0}^{4}a_{s}(\beta-1)^{s}\}\exp(-M\beta)\big]$} & {\tiny{}$B+\exp[-M(\beta-1)^{3}]\sum_{s=\text{1}}^{4}b_{s}(\beta-1)^{s}$} & {\tiny{}$C\beta^{2}\exp[\sum_{s=1}^{7}c_{s}(\beta-1)^{s}]$}\tabularnewline
\hline 
\hline 
{\footnotesize{}}%
\begin{tabular}{c}
{\footnotesize{}$\frac{Parameters}{Constants}$}\tabularnewline
{\footnotesize{}$M$}\tabularnewline
{\footnotesize{}$a_{0}$}\tabularnewline
{\footnotesize{}$a_{1}$}\tabularnewline
{\footnotesize{}$a_{2}$}\tabularnewline
{\footnotesize{}$a_{3}$}\tabularnewline
{\footnotesize{}$a_{4}$}\tabularnewline
{\footnotesize{}Corr.}\tabularnewline
\end{tabular} & {\footnotesize{}}%
\begin{tabular}{c|c}
\multicolumn{1}{c}{{\footnotesize{}$\alpha_{0}\simeq2$}} & {\footnotesize{}$\alpha_{2}\simeq1$}\tabularnewline
{\footnotesize{}0.880214} & {\footnotesize{}4.1814}\tabularnewline
{\footnotesize{}1.70465} & {\footnotesize{}0.0522961}\tabularnewline
{\footnotesize{}0.976489} & {\footnotesize{}-0.770667}\tabularnewline
{\footnotesize{}2.74286} & {\footnotesize{}2.78766}\tabularnewline
{\footnotesize{}-2.8097} & {\footnotesize{}-2.45835}\tabularnewline
{\footnotesize{}1.36045} & {\footnotesize{}-0.148637}\tabularnewline
\hline 
\multicolumn{1}{c}{\emph{\footnotesize{}0.963441}} & \emph{\footnotesize{}0.989417}\tabularnewline
\end{tabular} & {\footnotesize{}}%
\begin{tabular}{c|c}
\multicolumn{1}{c}{{\footnotesize{}$\frac{Parameter}{Constants}$}} & {\footnotesize{}$\hat{\alpha}_{3}$}\tabularnewline
{\footnotesize{}$M$} & {\footnotesize{}1.45629}\tabularnewline
{\footnotesize{}$B$} & {\footnotesize{}2.0503}\tabularnewline
{\footnotesize{}$b_{1}$} & {\footnotesize{}-9.16895}\tabularnewline
{\footnotesize{}$b_{2}$} & {\footnotesize{}30.8241}\tabularnewline
{\footnotesize{}$b_{3}$} & {\footnotesize{}-41.314}\tabularnewline
{\footnotesize{}$b_{4}$} & {\footnotesize{}22.1584}\tabularnewline
\hline 
\multicolumn{1}{c}{{\footnotesize{}Corr.}} & \emph{\footnotesize{}0.998267}\tabularnewline
\end{tabular} & {\footnotesize{}}%
\begin{tabular}{c|c}
\multicolumn{1}{c}{{\footnotesize{}$\frac{Parameter}{Constants}$}} & {\footnotesize{}$\gamma_{1}$}\tabularnewline
{\footnotesize{}$C$} & {\footnotesize{}0.730891}\tabularnewline
{\footnotesize{}$c_{1}$} & {\footnotesize{}3.50858}\tabularnewline
{\footnotesize{}$c_{2}$} & {\footnotesize{}-34.1656}\tabularnewline
{\footnotesize{}$c_{3}$} & {\footnotesize{}153.848}\tabularnewline
{\footnotesize{}$c_{4}$} & {\footnotesize{}-373.845}\tabularnewline
{\footnotesize{}$c_{5}$} & {\footnotesize{}507.468}\tabularnewline
{\footnotesize{}$c_{6}$} & {\footnotesize{}-359.883}\tabularnewline
{\footnotesize{}$c_{7}$} & {\footnotesize{}102.995}\tabularnewline
\hline 
\multicolumn{1}{c}{{\footnotesize{}Corr.}} & \emph{\footnotesize{}0.999860}\tabularnewline
\end{tabular}\tabularnewline
\hline 
\end{tabular}\protect\caption{\label{tab:Tab IIa-II}\emph{\footnotesize{}Optimization parameters
of the Eq. \ref{eq: 5} approximant, }{\footnotesize{}(}\emph{\footnotesize{}case
$\beta>1$}{\footnotesize{}): Formulas and their constants for $\alpha_{0}$,
$\alpha_{2}$, $\hat{\alpha}_{3}$ and $\gamma_{1}$.}}

\par\end{centering}

\end{table}

\subsubsection*{Squeezed instance, $\beta>1$}

Case $\beta>1$ is instead more difficult to adjust in the whole interval
of frequencies since no additional weight is possible to use. The
kink of $\log_{10}|\psi_{\beta}|$ near $\nu\sim1$ claims for a body
not overlooked which would be the case if tails were given more importance
by weighing them as in previous procedure. Besides, the mollifier
of Havriliak-Negami function is not enough elaborated and as a consequence
appears a bifurcation for each curve of parameters, corresponding
the lower branch to the best adjustment to data. Nevertheless if the
latter is employed for describing the curves, an abrupt change in
trend for them is evident and makes more difficult the handled mathematical
expressions in parameter adjustment. We show here only the upper branch
of all curves, this leads to a smooth and nice interpolation line
for each parameter as seen in tables \ref{tab:Tab IIa-II} and \ref{tab:Tab IIb-II}.
\begin{table}
\begin{centering}
\begin{tabular}{cc}
\hline 
{\tiny{}$\exp\big[(\frac{1}{\beta})^{p}\sum_{s=0}^{5}d_{s}(\beta-1)^{s}\big]$} & {\tiny{}$\exp\big[-M(\beta-1)^{0.1}\big]\sum_{s=1}^{4}q_{s}(b-1)^{s}$}\tabularnewline
\hline 
\hline 
{\tiny{}}%
\begin{tabular}{cr|r|r|c}
{\footnotesize{}$\frac{Parameters}{Constants}$} & \multicolumn{1}{r}{{\footnotesize{}$\tau_{0}$}} & \multicolumn{1}{r}{{\footnotesize{}$\tau_{1}$}} & \multicolumn{1}{r}{{\footnotesize{}$\tau_{2}$}} & {\footnotesize{}$\tau_{3}$}\tabularnewline
{\footnotesize{}$p$} & {\footnotesize{}$\equiv11.5$} & {\footnotesize{}$\equiv11.5$} & {\footnotesize{}$\equiv4.4$} & {\footnotesize{}$\equiv9.8$}\tabularnewline
{\footnotesize{}$d_{0}$} & {\footnotesize{}1.40526} & {\footnotesize{}1.91075} & {\footnotesize{}-0.0241828} & {\footnotesize{}1.36359}\tabularnewline
{\footnotesize{}$d_{1}$} & {\footnotesize{}-38.3857} & {\footnotesize{}-27.9873} & {\footnotesize{}-4.7472} & {\footnotesize{}-30.3907}\tabularnewline
{\footnotesize{}$d_{2}$} & {\footnotesize{}503.134} & {\footnotesize{}484.97} & {\footnotesize{}10.0218} & {\footnotesize{}380.42}\tabularnewline
{\footnotesize{}$d_{3}$} & {\footnotesize{}-3368.51} & {\footnotesize{}-2810.17} & {\footnotesize{}-31.8961} & {\footnotesize{}-1955.88}\tabularnewline
{\footnotesize{}$d_{4}$} & {\footnotesize{}8385.44} & {\footnotesize{}7209.62} & {\footnotesize{}44.4373} & {\footnotesize{}3128.49}\tabularnewline
{\footnotesize{}$d_{5}$} & {\footnotesize{}-9737.49} & {\footnotesize{}-7533.20} & {\footnotesize{}-28.0215} & {\footnotesize{}-2939.76}\tabularnewline
\cline{2-5} 
{\footnotesize{}Corr.} & \multicolumn{1}{r}{\emph{\footnotesize{}0.999708}} & \multicolumn{1}{r}{\emph{\footnotesize{}0.999695}} & \multicolumn{1}{r}{\emph{\footnotesize{}0.999214}} & \emph{\footnotesize{}0.999346}\tabularnewline
\end{tabular} & {\footnotesize{}}%
\begin{tabular}{c|c}
\multicolumn{1}{c}{{\footnotesize{}$\frac{Parameter}{Constants}$}} & {\footnotesize{}$\lambda$}\tabularnewline
{\footnotesize{}$M$} & {\footnotesize{}5.82578}\tabularnewline
{\footnotesize{}$q_{1}$} & {\footnotesize{}273.583}\tabularnewline
{\footnotesize{}$q_{2}$} & {\footnotesize{}-407.192}\tabularnewline
{\footnotesize{}$q_{3}$} & {\footnotesize{}-3.32942}\tabularnewline
{\footnotesize{}$q_{4}$} & {\footnotesize{}195.919}\tabularnewline
\hline 
\multicolumn{1}{c}{{\footnotesize{}Corr.}} & \emph{\footnotesize{}0.998083}\tabularnewline
\end{tabular}\tabularnewline
\hline 
\end{tabular}\protect\caption{\label{tab:Tab IIb-II}\emph{\footnotesize{}Optimization parameters
of the Eq. \ref{eq: 5} approximant, }{\footnotesize{}(}\emph{\footnotesize{}case
$\beta>1$}{\footnotesize{}): Formulas and their constants for $\tau_{0}$,
$\tau_{1}$, $\tau_{2}$, $\tau_{3}$ and $\lambda$. See figure \ref{Figure 2-II}.}}

\par\end{centering}

\end{table}

Although we started with a nine parameters \emph{ansatz }for the approximant
$\hat{\mathcal{AM}_{g,2}HN(\omega)}$ is clear from the graphs, (right
panel of figure \ref{Figure 2-II}), that $\alpha_{0}$ and $\alpha_{2}$
are almost constants. Now the written requirements over them are amply
fulfilled. Only there is a small disagree of order $10^{-4}$ from
condition $\alpha_{2}(\beta)\simeq1$ for some values of $\beta$.
This is entirely due to competition among parameters and subsequent
numerical errors. Meanwhile $\alpha_{0}(\beta)\approx2$ for all betas,
and any of them differ from this number less than 1.5\% for $\beta>1$
and only with some significance for beta 1.00 and 1.02, and for $\beta>1.80$.
Thus with a slight setting of $\hat{\mathcal{M}}_{g}(\omega)$ has
to be possible to write $\mathcal{AM}_{g,2}HN(\omega)$ as a seven
parameter function which is more economic computationally. (All the
adjustments to this set of parameters are given in tables \ref{tab:Tab IIa-II}
and \ref{tab:Tab IIb-II}).

\subsection{The role of the 'aide-de-camp' in the modified approximant}

Apart from already explained conditions in the onset of frequencies
which makes a Cole-Davidson relaxation suitable to describe the boundary
condition of $\frac{d\psi_{\beta}}{d\omega}$, it is obvious that
in the interval \textbf{$\beta\in(0.80,1.00]$} the first term of
approximant described in Eq. \ref{eq: 3} plays an important role
in the approximation since $\lambda$ is not at all negligible. However
for the interval $\beta\in[0.02,0.80]$ the story is quite different,
almost all its contribution is forced by theoretical considerations
as now the share coefficient is really small. To extend this situation
and make an adjustment with a one-term approximant in the whole interval
$\beta\in(0,1)$, we should prepare a more flexible second term of
Havriliak-Negami type in Eq. \ref{eq: 3}. And with this goal in mind
we establish the exponent $N$ of $\hat{\mathcal{M}_{l}}(\omega)$
as a new parameter of the optimization. Namely we shall do the following
setting:
\begin{equation}
\psi_{\text{\ensuremath{\beta}}}(\omega)\cong\mathcal{AM}_{l,1}HN(\omega)\equiv\frac{1}{(1+\mathcal{M}_{l}(\omega)(i\omega\tau_{2})^{\alpha_{2}})^{\frac{\beta}{\alpha_{2}}}},\label{eq: 7}
\end{equation}
with an estimator to $\mathcal{M}_{l}(\omega)$ similar to that of
Eq. \ref{eq: 4} though now $N=N(\beta)$ is non constant.

\begin{figure}
\centering{}\includegraphics[width=0.8\columnwidth]{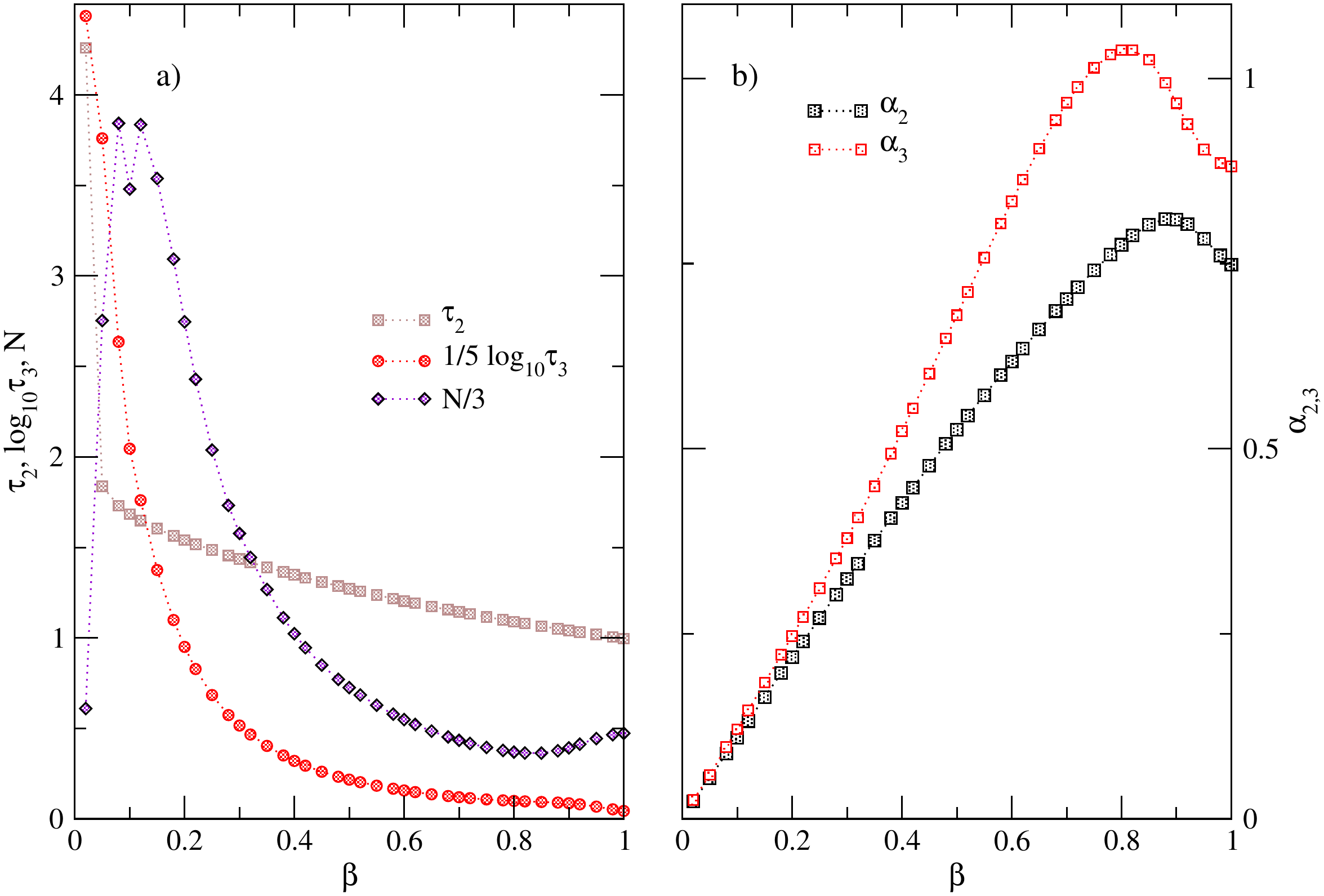}\protect\caption{{\footnotesize{}\label{Figure 3-II}Fitting parameters of Eq. \ref{eq: 7}
as functions of variable $\beta\leq1$. Left panel: characteristic
times $\text{\ensuremath{\tau}}_{2}$, $\tau_{3}$ and exponent $N(\beta)$,
(see Eq. \ref{eq: 4}). Right panel: frequency exponents $\alpha_{2}$
and $\alpha_{3}$, ($\hat{\alpha}_{3}$ in Eq. \ref{eq: 4}). The
dotted lines are just guides for the eye.}}
\end{figure}

The results, (\emph{i.e.} the parametric curves of $\beta$), are
shown in figure \ref{Figure 3-II}, there we note two important features
about the behaviour of parameters $\alpha_{2}$ and $\hat{\alpha}_{3}$
and the shape peculiarities of $N(\beta)$. The first characteristic,
in the interval $\beta\in(0.80,1.00]$, is that we do not recover
the functional form of a Debye relaxation as $\beta\rightarrow1$.
For such a requirement it should happen at least $N\rightarrow0$
and $\alpha_{2}\rightarrow1$ as an strong condition, or $\alpha_{2}+N\hat{\alpha}_{3}\simeq1$
as a weaker one, in that limit. Neither the strong nor weak conditions
are fulfilled by the parameters as can be seen in right and left panels
of figure \ref{Figure 3-II}.

In light of the share coefficient behaviour ($\lambda$) remains an
important question: if the auxiliary term dominated by it in the modified
approximant is really necessary. (See left panel in figure \ref{Figure 1-II}).
Or if instead it is only needed to 'unfreeze' the exponent $N$ in
the estimator $\hat{\mathcal{M}_{l}}(\omega)$ of the mollifier, (see
Eq. \ref{eq: 4}), to adjust $\psi_{\beta}(\omega)$ properly with
only one term: the mollified Havriliak-Negami function.

The second flaw is patent when we realize that it is not possible,
in the interval $\beta\in(0.02,0.10)$, to hold the condition $\alpha_{2}+N\hat{\alpha}_{3}\geq1$
when $\alpha_{2}\rightarrow0^{+}$ and $\hat{\alpha}_{3}\rightarrow0^{+}$
since $N$ is finite and decreasing as $\beta\rightarrow0$. These
trends of alpha parameters are attested, jointly with the $N$ one,
and depicted in figure \ref{Figure 3-II} again. In conclusion the
modified relaxation of Havriliak-Negami fails in the adjustment at
both ends of $\beta$ interval and it is not hard to imagine the difficulties
it has to describe an environment of $|\psi_{\beta}|\sim1$, (\emph{i.e.
$\omega\approx0$}), with a poor sampling of very low frequencies,
(as is the case of ours for so small values of beta). The tails obviously,
in such a situation, lead the adjustment and the mentioned requirements
about the behaviour of $\frac{d\psi_{\beta}}{d\omega}$ near $\omega\approx0$
should be imposed externally. At this event the best option to save
both flaws is to maintain the optimization with a two-terms approximant
like that of Eq. \ref{eq: 3}.

At last a further reason to keep the expression of Eq. \ref{eq: 3}
is to have a formal similarity with Eq. \ref{eq: 5} for the case
$\beta>1$ which makes more manageable the treatment and analysis
of the problem in the whole interval $\beta\in(0,2)$.

\section{Comparison and discussion: Suitability of formulae}

In light of these circumstances we will consider the frequency-averaged
relative error, (among data $\psi_{\beta}(\omega)$ and test functions),
as an indicator of reconstruction capability for any of the proposed
Havriliak-Negami approximants. As it has been patent till now most
of the present discussion here refers to the suitability of pairs
proposed to describe the modulus of data $|\psi_{\beta}|$. We must
be aware that aside from the results here discussed some additional
tuning of phase should be sought. Different one with each model for
approximation we use. Even so, without all the benefits of the phase,
an accurate adjustment between data and approximants makes this methodology
of multiple Havriliak-Negami summands useful to determine form parameters
in dielectric spectroscopy experiments, or in systematic search of
them by means of genetics algorithms.
\begin{figure}
\begin{centering}
\includegraphics[width=0.8\columnwidth]{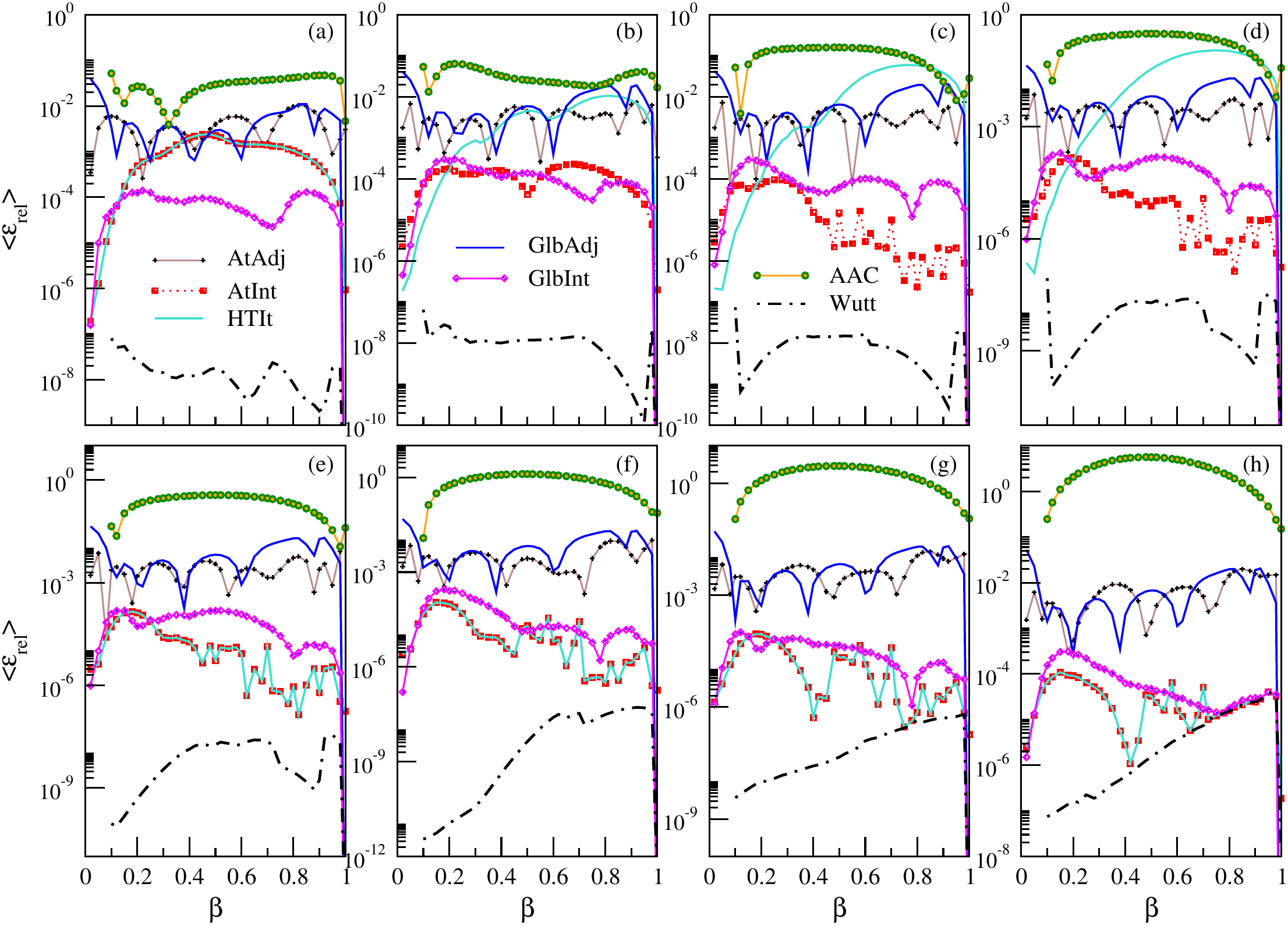}
\par\end{centering}

\centering{}\protect\caption{\label{Figure 4-II} {\footnotesize{}Frequency-averaged relative error,
$\langle\varepsilon_{rel}(\psi_{\beta}^{(a)},\psi_{\beta}^{(b)},\omega_{m},\omega_{x})\rangle$,
for data $\psi_{\beta}\equiv\psi_{\beta}^{(a)}$ obtained with Mathematica}\protect\textsuperscript{TM}{\footnotesize{}
and function values $\psi_{\beta}^{(b)}$ of models AAC (circle),
Wutt (dot-dashed), AtAdj (plusses), AtInt (squares), HTIt (light solid),
GlbAdj (dark solid) and GlbInt (diamonds). Eight frequency windows,
whose details are written in the text, and two types of samplings
--linear from (a) to (d) and logarithmic from (e) to (h)-- are shown
for interval $0<\beta\leq1$.}}
\end{figure}

Previously, a frequency-averaged relative error for the moduli of
functions was defined as:
\[
\langle\varepsilon_{rel}(\psi_{\beta}^{(a)},\psi_{\beta}^{(b)},\omega_{m},\omega_{x})\rangle=\frac{\int_{\omega_{m}}^{\omega_{x}}|1-\frac{|\psi_{\beta}^{(b)}|}{|\psi_{\beta}^{(a)}|}|\mathrm{d}\omega}{\omega_{x}-\omega_{m}}.
\]

Such error function is depicted in graphics \ref{Figure 4-II} and
\ref{Figure 5-II}. There it was calculated for the different models
described in a previous paper, \cite{Medi 2015}, and in the present
one, and also for other two models found in literature (see references
\cite{Wutt 2009,Alva 1991}). In particular, the errors for a Havriliak-Negami
function, the three models described above (double H-N, atlas of aproximants
and modified H-N), two variants of the last ones with parameters calculated
via formulae given in tables 1 to 4 in \cite{Medi 2015}, and the
numerical solution obtained from the \texttt{C} code in reference
\cite{Wutt 2009} are given, when $\beta\leq1$, in figure \ref{Figure 4-II}.
Same models with $\beta>1$, except 1HN of AAC (Ref. \cite{Alva 1991}),
are portrayed in figure \ref{Figure 5-II}.

The frequency windows studied are: (a) $[\omega_{m},\omega_{x})/2\pi$
$=[0,1)$, (b) $[1,10)$, (c) $[10,100)$, (d) $[100,500]$ in the
upper row of both figures. They followed a linear sampling with $\delta\nu$
$=\frac{1}{999.999}$, conversely the lower rows were logarithmic,
homogeneous in each decade in the way we already explained. Their
intervals of frequencies are: (e) $[100,10^{3})$, (f) $[10^{3},10^{6})$,
for both graphs, but (g) $[10^{6},10^{9})$, (h) $[10^{9},10^{12}]$
with $\beta\leq1$, and (g') $[10^{6},10^{7}]$ when $\beta>1$.

For a quick sight inside the plots in \ref{Figure 4-II} and \ref{Figure 5-II}
we have tagged the models already explained. We remind that $\psi_{\beta}^{(a)}(\omega)\equiv\psi_{\beta}(\omega)$
was obtained from the direct calculation of Fourier integral and is
the same reference for all errors $\langle\varepsilon_{rel}\rangle$
calculated with different test models $\psi_{\text{\ensuremath{\beta}}}^{(b)}$
which are now listed as: \emph{AAC}, the Havriliak-Negami approximation
cited in ref. \cite{Alva 1991}. \emph{Wutt,} the \texttt{C }library
of reference \cite{Wutt 2009} which employs the power series for
low and high frequencies and an effective numerical method for the
intermediary frequencies in the interval $\beta\in$ $[0.1,2.0]$.
\emph{AtAdj} is the label assigned to the model of equation \ref{eq: 2},
and the same is true for the symbol \emph{AtInt}. The distinction
is that while the parameters $\{\alpha_{1,2},\gamma_{1,2},\tau_{1,2},\lambda\}$
in the first case are calculated following the formulas in tables
1 to 8 of \cite{Medi 2015}, in the last one are obtained from the
points directly obtained of error minimization and depicted in graphics
5, 6, 7 and 8 of \cite{Medi 2015}. For the sake of clarity we have
repeated the latter results showing separately each part which follows
the eq. \ref{eq:1} for low or high frequencies (head and tail functions
of equation \ref{eq: 2}). It allows to appreciate where in the frequency
interval the individual approximant diverges from the data, and which
one is exactly its contribution to the atlas of approximants. The
transition from a plain downhill to a potential tail ($\nu^{-\beta}$)
it is then quite clear. This is called \emph{HTIt}. Besides all the
three previous models refer to the range $[0.02,2.00]$ of shape parameter
$\beta$.

The last assertion it is also true for labels \emph{GlbAdj} and \emph{GlbInt}
though two different formulas and their respective implementations
are employed, the equations \ref{eq: 3} and \ref{eq: 4} for $\beta\leq1$
and the equations \ref{eq: 5} and \ref{eq: 6} for $\beta>1$. Again
the first tag refers to the adjusted parameters (see tables \ref{tab:Tab Ia-II},
\ref{tab:Tab Ib-II}, \ref{tab:Tab IIa-II} and \ref{tab:Tab IIb-II})
and the second to the original points as depicted in figures \ref{Figure 1-II}
and \ref{Figure 2-II}.

\subsection{Models response in the stretched case}

\begin{figure}
\begin{centering}
\includegraphics[width=0.8\columnwidth]{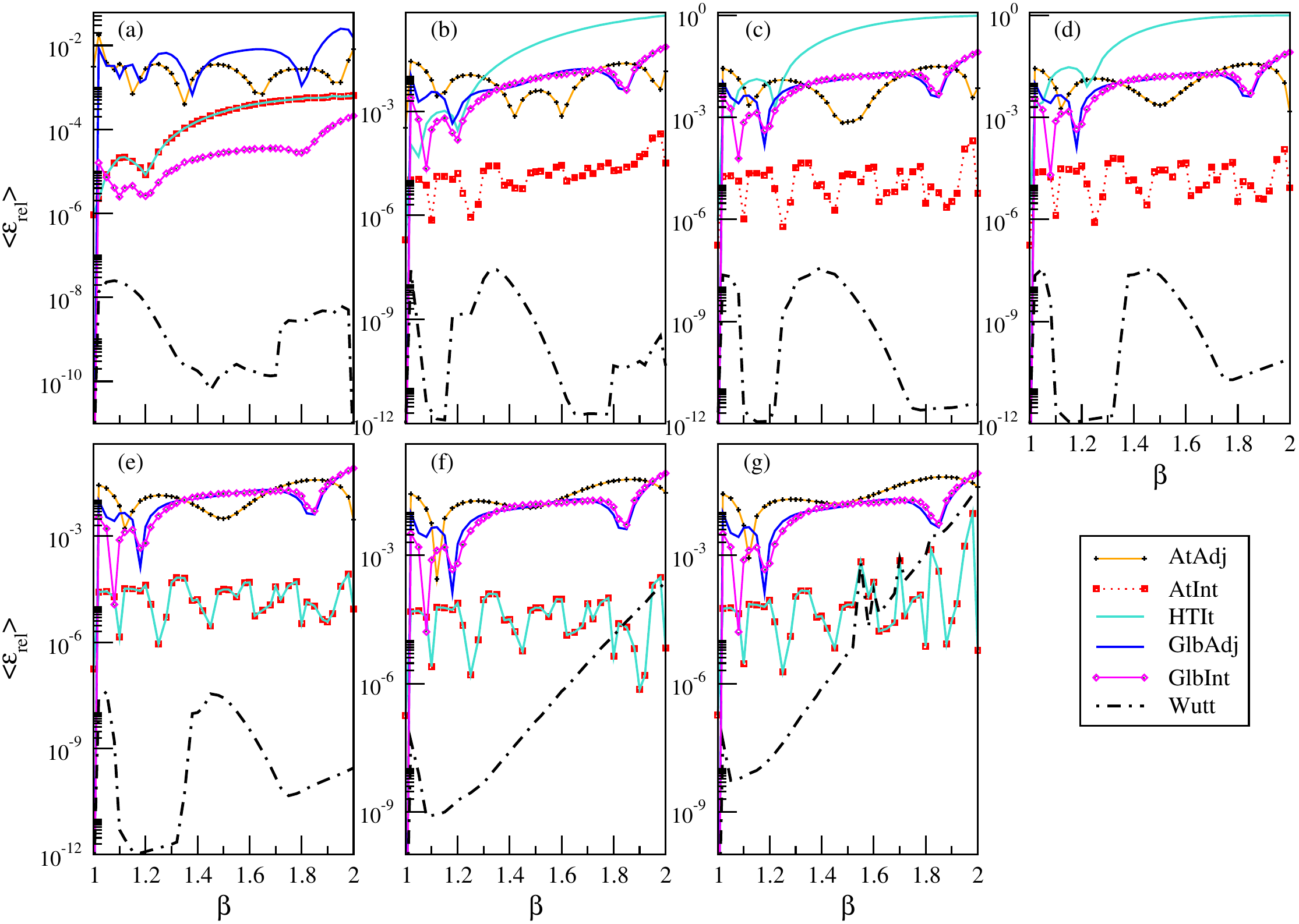}\protect\caption{\label{Figure 5-II}{\footnotesize{}Relative error, $\langle\varepsilon_{rel}(\psi_{\beta}^{(a)},\psi_{\beta}^{(b)},\omega_{m},\omega_{x})\rangle$,
for data $\psi_{\beta}\equiv\psi_{\beta}^{(a)}$ obtained with Mathematica}\protect\textsuperscript{TM}{\footnotesize{}
and test values $\psi_{\beta}^{(b)}$ of models: Wutt (dot-dashed),
AtAdj (plusses), AtInt (squares), HTIt (light solid), GlbAdj (dark
solid) and GlbInt (diamonds). Seven frequency windows and two types
of samplings --linear from (a) to (d) and logarithmic from (e) to
(g)-- are shown. Maximum frequency in panel (g) is reduced to $\nu_{x}=10^{7}$
due to glitches and numerical noise of data beyond this point. $1<\beta\leq2$.}}

\par\end{centering}

\end{figure}

In figure \ref{Figure 4-II},\emph{ AAC}, the approximation with only
one HN function, shows the biggest of all errors for the models here
presented and the interval $0<\beta\leq1$. The best result is of
course that of \emph{Wutt} which combine analytics and numerics. Meanwhile
the atlas described by equation \ref{eq: 2}, (\emph{i.}e. \emph{AtInt}),
works quite well even in the interval of very low frequencies where
we demonstrated that one of the functions of the approximant should
be a Cole-Davidson relaxation or a modified version of it, --this
depends on if $\beta$ is less than or greater than unit. (See Eqs.
\ref{eq: 3} and \ref{eq: 5}). Besides, it holds good terms in the
medium range thanks to the change of describer function, (\emph{i.e.
}from head, $\mathcal{A}p_{2,l}HN$, to tail, $\mathcal{A}p_{2,h}HN$;
see figure \ref{Figure 4-II} panel (b)). The performance of this
swap is even better for high frequencies because the matching with
data exceeds expectations and the approximation has not required any
restriction in the product $(\alpha\cdot\gamma)_{1,2}$. This reinforces
our previous conclusion \cite{Medi 2011}, and points to the true
nature of $\psi_{\beta}$ as a sum of a Havriliak-Negami pair with
almost constant coefficients which only change significatively as
$1>\omega\rightarrow0$ \cite{Weis 1994}. Such important transition
is highlighted in figures \ref{Figure 4-II} and \ref{Figure 5-II}
by the discrepancy between model \emph{AtInt} (squares) and \emph{HTIt}
(light solid line) and reminds us the need for at least two charts
(one for $\nu<1$ the other for $\nu>1$) in the description of the
whole function $\psi_{\beta}$. Usually this approach is not taken
in consideration in the literature since only one set of parameters
$\{\alpha,\gamma,\tau,\lambda\}$ is employed to match the data, or
if considered is misinterpreted due to the usage of a time scale factor
in the stretched exponential (\emph{i.e. }$\exp-(t/\tau_{KWW})^{\beta}$)
\cite{Schr 2010}. See panels (a) to (d) in figure \ref{Figure 4-II}.

It is worth to note how the Havriliak-Negami approach \emph{AAC} starts
to work better than the double approximant \emph{HTIt} in the range
of medium-to-high frequencies, according as $0.8<\beta\rightarrow1$.
It sounds logical since $\text{\ensuremath{\beta}}\rightarrow1$ implies
$\lambda\rightarrow0$, and this final value is a pathology for the
double sum of HN functions quite difficult to treat numerically. Such
problem is not evident using the atlas \emph{AtInt }since the approximant
of tail\emph{ $\mathcal{A}p_{2,h}HN$ }takes the control over $\mathcal{A}p_{2,l}HN$
in such frequency interval. The former function besides, at very high
$\nu$'s and with $\beta$ near $1.00$, shows similar errors to the
results of numerical-analytic method \emph{Wutt}. Now the major problem
for both of them will be the numerical oscillations of reference data.
See panels (g) and (h) in figure \ref{Figure 4-II}. 

The differences of error between model \emph{AtAdj} and \emph{AtInt}
present clearly two regions. One in the low frequencies zone ($\nu<1$)
the other in the high frequency one ($\nu>1$), as it is usual coinciding
with the onset of potential behaviour for tails. In the first case
it is the lack of ability of the double approximant with constant
parameters to approach data, what gets closer both models. This is
shown in figure \ref{Figure 4-II}a and is more conspicuous when $\beta\sim0.40$.
In the second case where both models split apart more than one order
of magnitude the reason for this is more subtle because the correspondence
between $\psi_{\beta}^{(a)}$ and $\psi_{\beta}^{(b)}$ is tighter
as both functions follow a quite similar potential decaying. (See
graph \ref{Figure 4-II}, panels (b) to (h), and figure 9 in \cite{Medi 2015}).
What makes such difference between \emph{AtInt} and \emph{AtAdj} is
an extra error provided for each of the parameters $\alpha_{1,2}$,
$\gamma_{1,2}$, $\tau_{1,2}$ and $\lambda$. An additional contribution
which is consequence of the fitting of parametric curves to optimized
points. Then it would be desirable to reduce those inputs binding
the parameters to relationships as the already mentioned $(\alpha\cdot\gamma)_{1,2}\sim\beta$.
Nevertheless, we feel that some further work should be done to link
the conduct of $\tau(\text{\ensuremath{\beta}})_{1,2}^{\beta}$ and
$\lambda$ to the coefficients of the analytical series for $\psi_{\beta}$,
and so in the absence of them we have presented the models drawn from
eq. \ref{eq: 2} free of any external constraints.

As a novelty we introduced here two global models to simulate the
data, namely\emph{ GblInt} (optimized points) and \emph{GblAdj} (adjusted
parameters). Watching them carefully in the various intervals of frequency
one can realize how the performances are similar to the models of
Havriliak-Negami double approximants, \emph{AtInt} and \emph{AtAdj},
respectively. Also is possible to observe how for the low frequency
range, $\nu<1$, the model \emph{GblInt }outperforms to\emph{ AtInt},
although this one later improves and is usually better for the high
frequencies in accuracy. (See in panels (a) to (h) of figure \ref{Figure 4-II}
the square and diamond curves). Also for the global model, the aggregated
error of several independent parameters spoils the result making alike
the error curves of cases \emph{AtAdj} and \emph{GblAdj}, (plusses
and dark solid lines in graphics). Therefore the drastic displacement
of relative error lines towards a lesser precision for models with
adjusted parameters $\{\alpha_{1,2},\gamma_{1,2},\tau_{1,2},\lambda\}(\beta)$
points to the need for strict relationships among them and good descriptions
of functional dependence with the shape parameter $\beta$.

However it is important to emphasize how proposed estimators of mollifiers,
($\hat{\mathcal{M}}_{l,g}$ in Eqs. \ref{eq: 4} and \ref{eq: 6}),
are subject to many ``\emph{ad hoc}'' restrictions, mainly deduced
of data traits and information obtained from the behavioral changes
of $\alpha_{1,2},\gamma_{1,2},\tau_{1,2},\lambda$ curves while changing
the regime of frequencies from low to high. The most significative
restraint here is the fact the $\hat{\mathcal{M}}_{l,g}$ functions
are set as real ones when they should value in the complex field.
As we have seen before, in Eqs. \ref{eq: 3} and \ref{eq: 5}, near
$\omega\approx0^{+}$ the dominant demeanor of $\hat{\mathcal{M}}_{l,g}(\omega)$
is determined without consider further modifications to the phase
$i^{\alpha_{2}}$ since the frequency factor $\omega^{\alpha_{2}+*}$
extinguishes such contribution quickly and only remains the one of
extended Cole-Davidson term. Nevertheless the role the phase plays
is over the entire interval of frequencies and, although to our present
purpose of describing the modulus accurately it is not crucial this
bias in the argument of the approximation $\mathcal{AM}_{l,g,2}HN$,
it is important to point out the need of a mollifier in complex series
for proper description of $\psi_{\beta}$.

It seems promising, from an analytical point of view, that a strategy
to sum up a quite difficult series in the neighborhood of $\omega\approx0^{+}$
comes from the help of an extended Havriliak-Negami pair. Perhaps
could be interesting to decide the mollifier's functional form with
the aid of series, integrals or equations which determine $\psi_{\beta}$.
Mainly when $\omega\rightarrow0$, for $\beta<1$, or when $\omega\rightarrow\infty$
for $\beta>1$, the most difficult cases for power series involved
\cite{Wint 1941,Dish 1985,Schr 2010}.

Obviously, in the light of the problems we face when using unsuitable
functions as estimators of $\mathcal{M}_{l,g}(\omega)$, it is suggested
that a similar 'loss' of phase could happen in the atlas approximation
of eq. \ref{eq: 2} to data. And consequently foresee a mild shift
in the argument of whole function $\mathcal{A}p_{2}HN(\omega)$ at
high frequencies. This displacement is due to the way the tail functions
$\mathcal{A}p_{2,h}HN(\omega)$ are determined. Data are pruned in
a logarithmic pace and the important information at low frequencies,
--the plateau--, is removed when optimizing tail parameters, something
is not made in the case of head functions, $\mathcal{A}p_{2,l}HN(\omega)$.
All this does not impact very much on the approach to the modulus,
as we saw in graphics of figures \ref{Figure 4-II} and \ref{Figure 5-II},
but suggests a share parameter fully complex, \emph{i.e. $\lambda\in\mathbb{C}-\mathbb{R}$}.
Unfortunately it would add a new degree of freedom and would overshadow
the discussion about modulus characteristics in absence of a thorough
treatment of the data phase.

Apart its mentioned inability to describe the very low frequency range,
the double Havriliak-Negami approximant, $\mathcal{A}p_{2}HN_{\alpha,\gamma,\tau,\lambda}(\omega)$,
will not present such difficulties while describing the argument,
--much less the modulus--, of data $\psi_{\beta}$.

\subsection{Models response in the squeezed case}

The last stay in this discussion is the figure \ref{Figure 5-II}
that shows relative errors of the six previous models for shape parameter
interval $1<\beta\leq2$. As predicted by first and second logarithmic
derivative, (see frames a) and b) of fig. 2 in \cite{Medi 2015}),
there is a sudden change of behaviour in the slope of $|\psi_{\beta}|$
from flatness to a potential decline $\omega^{-\beta}$ in a relatively
small interval of frequencies. (See frame c) of same graphics in \cite{Medi 2015}).
This is a much more sharper and distinct transition than in case $\beta\leq1$,
which forces the existence of a different set of constraints for products
$\alpha\cdot\gamma$ in the double Hav.-Neg. approximation, as clearly
shows figure 9 in \cite{Medi 2015}. All this oblige to abandon quickly
the \emph{HTIt} model\emph{,} (light solid line in \ref{Figure 5-II},
panels (a) to (d)), in favor of \emph{AtInt }because the latter holds
itself quite close to the potential tail and shares same description
of $|\psi_{\beta}|$ with the former at very low frequencies. (See
squares inside panels (a), and (b) to (g) in figure \ref{Figure 5-II}).

Again, as with $\beta\leq1$, a big distance in terms of relative
error separates \emph{AtInt} and \emph{AtAdj}, and as before this
gap is attributed to a collective error subscribed by each individual
parameter, whenever every uncertainty is caused by obtaining the appropriate
parameter from a pertinent fitting function along all values of $\beta$.
However for the models \emph{GlbInt} and \emph{GlbAdj} such distance
doesn't exist at medium frequencies and thereafter, \emph{i.e. }$\nu>1$.
(See diamonds and dark solid line in graphics of figure \ref{Figure 5-II}).
It seems that model \emph{GlbInt} it is not able to keep track of
data tail so close as \emph{AtInt }does. Surely the 'bi-chronicity'
or the mollifier in functional form of Eq. \ref{eq: 5} should be
revisited to give a proper account of directional twist of data near
$\nu\sim1$, and thus to diminish the error below the collective contribution
of parameters. As in fact it happens at low frequencies, (see figure
\ref{Figure 5-II}a). Nevertheless, as far we know, this is one of
the few attempts to describe globally the Fourier transform $\psi_{\beta}$
for $\beta>1$ with an analytical model albeit approximate, so each
piece of formula $\mathcal{AM}_{g,2}HN(\omega)$ has great value for
future mathematical inquiries.

Finally we should point out how, at very large frequencies, the potential
description overcomes some numerical difficulties experienced by \emph{Wutt}
model for large $\beta$ values, \emph{i.e.} $\beta\geq1.60$. (See
figures \ref{Figure 5-II}f and \ref{Figure 5-II}g).

\section{Conclusions}

The present work is devoted to a compact description of the Fourier
Transform of the Kohlrausch relaxation. As any reconstruction of this
function as from spectral data should heavily depend on the information
of frequencies near zero since tails will be surely corrupted by noise,
an extra effort has to be made to manufacture a global function mimicking
all aspects of this transform from low to high frequencies. Thus,
two new sets of such models are proposed, and a detailed discussion
on their errors compared to numerical control calculations generated
directly by evaluating the Fourier integrals is made. 

We found how the approximation with a double H-N function always underestimates
$\psi_{\beta}$ in the low frequency range, ($\nu<0.1$). And although
it is a small difference forces us to change those values of parameters
$\{(\alpha,\gamma,\tau)_{1,2},\lambda\}$ already obtained in the
range of medium frequencies.

This repeats again in the transition from medium to high, or very
high, frequencies. Nevertheless the double Havriliak-Negami approximation
is close enough to the original function as to describe it along a
wide range of frequencies before the variation can be noticed. Moreover
the parameters should not be regarded as varying, if the interval
where the approximation is performed only comprises very high frequencies.

Thus, due to the slow variation of parameters of the approximant with
the frequency, (\emph{adiabatic} \emph{parameters}), instead of a
global function with $\omega-$dependent $\{(\alpha,\gamma,\tau)_{1,2},\lambda\}$,
we employed different charts of double Havriliak-Negami sums to describe
locally $\psi_{\beta}(\omega)$. We found that using two charts is
a good approximation, enough to establish an atlas, however the differences
at low frequencies still persist with such number of maps. The inclusion
of a third one in the neighborhood of zero, \emph{i.e. }$\nu\approx0$,
should be convenient, though the existence of an exact analytical
series of powers in terms of Cole-Cole relaxations \cite{Cole 1941,Weis 1994},
(\emph{i.e.} $1/(1+(i\omega)^{\beta})$), for $\psi_{\beta}$ suggests
the radius of such chart will depend on $\beta$ \cite{Schr 2010}.
What it makes difficult and cumbersome working with an atlas of three
maps.

The question whether it is possible to sum up the series of Cole-Cole
terms at $\nu=0$, or it is possible to write the atlas of Havriliak-Negami
charts just in a global way, seems to have a positive answer. We presented
two \emph{ansätze} for extending the double Havriliak-Negami approximation,
-- that has proved to be successful locally --, which describe along
several decades in frequency, and with enough functional proximity,
the data of $|\psi_{\beta}|$. By means of rough estimates of the
\emph{mollifiers} of these new testing relaxations, we have found
a good agreement with data moduli signaling a path for a future close
approximation in the complex domain to the series, or numerical integrals,
of the Fourier Transform $\psi_{\beta}$.

\end{document}